# Impact of Coastal Hazards on Residents' Spatial Accessibility to Health Services[1]

Georgios P. Balomenos, A.M.ASCE[1]; Yujie Hu[2]; Jamie E. Padgett, A.M.ASCE[3] and Kyle Shelton[4]

[1]Postdoctoral Fellow, Dept. of Civil and Environmental Engineering, Rice Univ., 6100 Main St., Houston, TX 77005 (corresponding author). E-mail: georgios.p.balomenos@rice.edu

[2]Assistant Professor, Dept. of Geography, Univ. of Florida, 3141 Turlington Hall, Gainesville, FL 32611; formerly, Assistant Professor, School of Geosciences, Univ. of South Florida, 4202 E. Fowler Ave., Tampa, FL 33620. Email: yujiehu@ufl.edu

[3]Associate Professor, Dept. of Civil and Environmental Engineering, Rice Univ., 6100 Main St., Houston, TX 77005. E-mail: jamie.padgett@rice.edu

[4]Director of Strategic Partnerships, Kinder Institute for Urban Research, Rice Univ., 6100 Main St., Houston, TX 77005. E-mail: kyle.k.shelton@rice.edu

**Abstract:** The mobility of residents and their access to essential services can be highly affected by transportation network closures that occur during and after coastal hazard events. Few studies have used geographic information systems coupled with infrastructure vulnerability models to explore how spatial accessibility to goods and services shifts after a hurricane. Models that explore spatial accessibility to health services are particularly lacking. This study provides a framework to examine how the disruption of transportation networks during and after a hurricane can impact a resident's ability to access health services over time. Two different bridge closure conditions—inundation and structural failure—along with roadway inundation are used to quantify post-hurricane accessibility at short- and long-term temporal scales. Inundation may close a bridge for hours or days, but a structural failure may close a route for weeks or months. Both forms of closure are incorporated using probabilistic vulnerability models coupled with GIS-based models to assess spatial accessibility in the aftermath of a coastal hazard. Harris County, an area in Southeastern Texas prone to coastal hazards, is used as a case study. The results indicate changes in the accessibility scores of specific areas depending on the temporal scale of interest and intensity of the hazard scenario. Sociodemographic indicators are also examined for the study region, revealing the populations most likely to suffer from lack of accessibility. Overall, the presented framework helps to understand how both short-

---

term functionality loss and long-term damage affect access to critical services such as health care after a hazard. This information, in turn, can shape decisions about future mitigation and planning efforts, while the presented framework can be expanded to other hazard-prone areas.

## Introduction

The transportation network is vital to ensure connectivity in areas damaged by natural hazards such as earthquakes, floods and hurricanes (Nicholson and Du 1997). In the event of an earthquake, several studies have already examined the consequences of road network closures as presented in Tamima and Chouinard (2017). However, few studies have focused on bridge and road closures in the event of hurricanes. Instead, many consider the transportation network as fully operational for supporting post-hurricane efforts (Horner and Downs 2008); while others limit their analyses to identifying infrastructure that falls within the inundation zone, without considering physical vulnerability, functionality loss or post-hurricane accessibility (Sohn 2006; Sánchez-Silva and Rosowsky 2008; Gomez et al. 2013). Closing this gap is essential because increasing numbers of people live in vulnerable coastal areas (Widener and Horner 2011) and the increase in extreme weather events (Kermanshah and Derrible 2017) and rising sea levels (Nicholls et al. 1999) have put road networks under constant threat of flooding and/or structural failure.

While authorities may issue mandatory evacuation orders prior to hurricane landfall, personal/socioeconomic issues may prevent smooth evacuation from affected regions (Horner and Widener 2011; Widener et al. 2013), e.g., evacuation can be difficult for vulnerable populations such as the elderly and people with a low income, among others (Shirley et al. 2012). In addition, in the aftermath of a hurricane there is an immediate need for those affected to access not just basic needs like food and water (Horner and Widener 2011), but also health services to address both chronic and emergency health issues (Ford et al. 2006; Bethel et al. 2011). This need is especially acute for vulnerable populations such as pregnant women and infants (Callaghan et al. 2007), populations prone to pre-existing health conditions, e.g., elderly people (Horner et al. 2018), and populations prone to hurricane damage and injuries, e.g., low-income people living in houses susceptible to hurricane damage (Bjarnadottir et al. 2011). In addition, recent studies have highlighted the importance of post-hurricane accessibility to health care facilities for people over 65 years of age and/or with low incomes (Behr and Diaz 2013; McQuade et al. 2018).

Locations of vulnerable populations could be identified by examining the service area of health care services (i.e., looking for those populations living outside of the service area). The delineation of health care service areas has been well studied in health care and geography research. One simple approach is to draw a circle around a service location using a threshold distance such as 8 km. The delineated circle constitutes the service area of the health care facility. However, this method assumes that people travel in a straight-line fashion without considering the road network layout. Therefore, some research proposed network-based distance measure for improved estimates (Hu et al. 2010); for example, all streets that can be accessible within one hour from a hospital form the service area of the hospital. Both methods only emphasize the spatial perspective while lacking consideration of other factors such as quality and costs of service in a health care location. This could be mitigated by analyzing actual service visit data. For example, the Dartmouth Atlas of Health Care (http://www.dartmouthatlas.org/) defined a nationwide Hospital Service Areas (HSA) and Hospital Referral Regions (HRR) based on the 1992-1993 Medicare hospitalization records. For instance, the HSA are defined by assigning each zip code to the city containing the hospitals most often used by local residents in the zip code. The method was also used to delineate health care service regions in other countries such as Switzerland (Klauss et al. 2005). However, Hu et al. (2018) argued that this approach involves randomness and arbitrary choices in the calculation and hence the final service areas are not reliable. Accordingly, they developed an optimization approach to define hospital service areas with the maximal within-area patient-to-hospital flow and minimal inter-area patient-to-hospital flow (Hu et al. 2018).

      A hazard event such as flooding can cause damage to the road network and limit people's mobility including their access to health care services (Taylor et al. 2006). In addition to the road network, the failure of other types of infrastructures such as a health care facility itself (e.g., lack of supply of water and electricity) can also play a role (Arboleda et al. 2009). Therefore, having an operational infrastructure network after a hazard event is of paramount importance. Among many types of infrastructures, this study is specifically interested in the transportation network including streets and bridges. Existing studies have already highlighted the structural vulnerability of bridges when exposed to coastal hazards. Hurricanes Ivan (2004) (Douglass et al. 2004), Katrina (2005) (Padgett et al. 2008) and Ike (2008) (Stearns and Padgett 2011) highlighted bridge failures where primarily the bridge deck was shifted or displaced due to uplift forces caused by storm surge and waves during these events. For example, Hurricanes Katrina and Ike caused damages to 44 and 26 bridges, respectively (Padgett et al. 2008; Stearns and Padgett 2011), with Hurricane Katrina revealing that bridges where the most vulnerable component of the transportation network (Padgett et al. 2012). During coastal hazard events bridges may experience significant structural failures, in

addition to inundation, which can affect network functionality and post-hurricane accessibility (Ataei and Padgett 2012; Kameshwar and Padgett 2014).

This study addresses current gaps in the literature by providing a framework that couples infrastructure inundation and vulnerability modeling from coastal storms with GIS-based spatial accessibility analysis. This framework is used to examine how the disruption of transportation networks due to bridge and roadway closure during and after a hurricane can impact a resident's ability to access health services over time. Along with roadway inundation, two bridge closure conditions (i.e. bridge inundation and bridge structural failure) are considered in the framework to provide insight on post-hurricane spatial accessibility to health services for two time horizons (i.e. short- and long-term). An application of the framework is provided for Harris County, Texas, under two different storm scenarios representing moderate and high intensity events. The results provide insights on the variation in spatial accessibility based on different temporal scales of interest and intensity of hazard. Furthermore, to provide context regarding affected populations, sociodemographic characteristics of groups with limited access to health care facilities are analyzed and discussed.

**Infrastructure Vulnerability Modeling**

In order to evaluate post-hurricane accessibility to health services, as described above, models that evaluate the vulnerability of bridges and roadway infrastructure and its implication on transportation network functionality are required. Two closure conditions are considered for bridges in this analysis. A bridge may be closed to traffic either because it is inundated or because it is structurally damaged. For roadways, although it is acknowledged that hurricanes may also lead to physical damage associated with washout and undermining, this study focuses on roadway inundation given its predominance in affecting post-hazard accessibility and given the lack of any existing vulnerability models on storm-induced roadway damage.

First bridge and roadway inundation is evaluated for the surge produced by the scenario storm of interest. For the case of bridge closure due to inundation, the metric of relative surge elevation is introduced which is calculated as (Fig. 1)

$$Z_c = H_b - H_{st} \tag{1}$$

where $Z_c$ is the relative surge elevation, i.e., negative if the storm surge level is above bridge deck, $H_b$ is the bridge deck height, and $H_{st}$ is the storm surge height. In this study, each bridge deck elevation is determined using LiDAR data (H-GAC 2017). Storm surge elevation is determined using the output from storm simulations provided by the SSPEED Center (SSPEED 2017) that were performed with an ADCIRC (ADvanced CIRCulation) model in conjunction with the Simulating Waves Nearshore (SWAN) model (Luettich and Westerink 2004). ArcGIS (ArcGIS 2013) is then used to calculate the relative surge elevation at each bridge location, where both the bridge deck and storm surge elevation are referenced to the same datum. When $Z_c$ is negative, the bridge deck is considered inundated and a safety limit is then imposed to map inundation to traffic restrictions. In most cases, bridges and roadways are considered inaccessible when the water level reaches a safety threshold of about 0.6 m (Anarde et al. 2017). Therefore, a bridge is considered closed due to inundation when $Z_c \leq -0.6$ m. Similarly, road closure due to flooding is calculated as (Fig. 2)

$$d_{in} = H_{st} - H_r \qquad (2)$$

where $d_{in}$ is the inundation depth, $H_r$ is the road height determined using LiDAR data (H-GAC 2017), and the road is considered closed due to inundation when $d_{in} \geq 0.6$ m.

The second closure condition modeled for bridges is that associated with structural failure. This study examines bridge uplift caused by hurricane-induced storm surge and waves, which is a common failure mode observed during past events (Padgett et al. 2008; Stearns and Padgett 2011). Ataei and Padgett (2012) developed fragility models for examining the uplift probability of the bridge deck conditioned on relative surge elevation and wave height assuming limited vertical connectivity between the deck and pier as typically observed along much of the US Gulf Coast. Fig. 3 shows the conditional probability of uplift for bridges with mean span mass per unit length of 15-20 ton/m. Furthermore, results by Ataei and Padgett (2012) reveal that of various classification metrics, the bridge mean span mass per unit length is the most appropriate metric for classifying vulnerability of bridges with different structural characteristics. Fig. 4 shows the uplift probability for different bounds of mean span mass, showing a significant reduction in uplift fragility with an increase of this parameter. The associated parameterized fragility model developed by Ataei and Padgett (2012) is adopted in this study and the probability of a bridge experiencing uplift is calculated as

$$p_f = a + bH_{max} + cZ_c \qquad (3)$$

where $p_f$ is the probability of failure, i.e., bridge deck unseating, $a, b, c$ are regression coefficients adopted from Ataei and Padgett (2012) based on the mean span mass per unit length for each examined bridge (Table 1), and $H_{max}$ is the maximum wave height estimated using AASHTO (2008) as

$$H_{max} = 1.8\, H_s \qquad (4)$$

where $H_s$ is the significant wave height. In this study, similar to $H_{st}$, $H_s$ values are provided by the SSPEED Center (SSPEED 2017), i.e., ADCIRC and SWAN are coupled to determine $H_s$ values for a study region for each examined storm, and then ArcGIS (ArcGIS 2013) is used to extract $H_s$ values at each bridge location. Given that $p_f$ by itself does not provide a deterministic threshold, but rather a probability of damage, contrary to the previous inundation closure condition, N numbers of samples are generated for each bridge to examine the structural failure closure condition. Thus, N numbers of samples with zeros and ones are populated for each bridge based on the calculated $p_f$, where one denotes bridge closure and zero denotes safe pass. In this way, structural failure conditions for each bridge are determined for a subsequent Monte Carlo simulation (MCS) network analysis.

**Spatial Accessibility Modeling**

Spatial accessibility defines the relative ease by which activities or services (in this case healthcare services) can be reached from a given demand location (such as a resident's home) based on spatial barriers such as the distance between them (Penchansky and Thomas 1981; Ikram et al. 2015). It measures the potential utilization patterns of services based on "where you live," instead of the actual utilization patterns (e.g., frequency of visits). Assessing residents' spatial accessibility can help planners and policymakers identify possibly underserved areas and target future plans to address those gaps.

Various models have been developed to measure spatial accessibility and applied in different disciplines such as transportation and public health. A commonly used accessibility measure is the proximal area method, which calculates the minimum travel cost (in distance or time) from a demand location to the location of the closest supply facility. For example, Onega et al. (2008) measured residents' spatial accessibility to cancer care services based on the network travel time to the nearest cancer care facilities. In addition to cancer care, other researchers have examined spatial accessibility to general practitioners, cancer screening facilities, psychiatric treatment and pharmacies using the same proxmial area method (Brabyn and Gower 2003; Wang et al. 2008; Mennis et al. 2012; Ikram et al. 2015). Though simple, this method may not accurately capture the accessibility concept as it assumes that residents only visit their closest supply locations. Other researchers have developed accessibility measures based on cumulative opportunites within a geographic range to include all possible supply locations but not the nearest. For example, Scott and Horner (2008) calculated accessibility as the number of a variety of opportunities including retail, service, leisure and religious facilities that are reachable within a given travel time. Owen et al. (2016) applied this method to measure accessibility to jobs. A variation of the cumulative opportunity metric is the gravity-based model, which further considers distance decay behavior—the idea that the farther the supply facility is located, the less likely a resident is to visit. The gravity measure of accessibility was first developed by Hansen (1959) and is frequently used in transportation studies (Thill and Kim 2005; Boisjoly and El-Geneidy 2016; Moya-Gómez and García-Palomares 2017). Despite incorporating more than the nearest opportunies in the calculation, these methods only target the supply side, neglecting demand (Ikram et al. 2015).

A more accurate and reliable accessibility measure, therefore, should consider both supply and demand as well as any spatial barriers between them. This is particularly relavent for post-hurricane accessibility to health services where there may be closures because of bridge damage or roadway inundation. The popular two-step floating catchment area (2SFCA) method, developed by Luo and Wang (2003), is one such approach. In the first step, the 2SFCA method simply calculates the supply-to-demand ratio $R_j$ within a catchment area—i.e., a distance range $d_0$—around each supply facility location j as

$$R_j = \frac{S_j}{\sum_{k\in\{d_{kj}\leq d_0\}} D_k} \quad (5)$$

where $S_j$ denotes the capacity of supply at the facility j, and $D_k$ represents the demand size at location k that lies within the catchment area (or service area) $d_0$ around facility j. However, some demand locations could be involved in the catchment areas associated with multiple supply facilities; in other words, there is more than one opportunity for the people at those demand locations and a simple supply-to-demand ratio $R_j$ cannot fully capture the accessibility of this population. To solve that issue a second step is followed to sum up the ratio $R_j$ at any supply facility location j that falls within the same catchment range from a demand location i and the result is the accessibility to services at demand location i. The second step is particularly designed for demand locations where residents have access to multiple supply locations. In fact, if a demand location i is only accessible to one supply location j, then the accessibility at location i is simply $R_j$; if residents living at demand location i have access to supply locations j and k, then the accessibility of demand i is $R_j + R_k$. Although not considered in this study, this adopted measure is particulary attractive in a post-hazard analysis since the damage or availability of the critical services or health care facility could also be incorporated as a reduction on the supply side. Let $A_i$ represent the accessibility at demand location i, it can be compuated as

$$A_i = \sum_{j\in\{d_{ij}\leq d_0\}} R_j \quad (6)$$

Substituting $R_j$ into Eq. (6) using Eq. (5), $A_i$ is then transformed as

$$A_i = \sum_{j\in\{d_{ij}\leq d_0\}} \frac{S_j}{\sum_{k\in\{d_{kj}\leq d_0\}} D_k} \quad (7)$$

A larger value of $A_i$ indicates a better accessibility at a location. One may refer to Luo and Wang (2003) for more technical details.

**Integrated Framework and Case Study**

Infrastructure vulnerability modeling and spatial accessibility modeling are integrated into a framework that can provide information on spatial accessibility to critical services after a hazard event. Fig. 5 provides a flowchart of the integrated framework which also examines the road inundation for each bridge closure condition. The proposed approach (Fig. 5) initializes with a scenario hazard event, i.e., requires storm characteristics as inputs (e.g., surge elevation and wave height) for calculating road and bridge closures; it is illustrated through a case study which examines residents' accessibility to health services to the east part of Harris County in Texas, i.e., an area prone to hurricanes. Two storm scenarios are applied for each bridge closure condition, while the results for each storm scenario are presented in terms of two time horizons, i.e., short-term effects and long-term effects. Short-term effects can refer to limited accessibility for hours or days due to bridge structural failure, bridge inundation, and road inundation. Long-term effects can refer to limited accessibility for weeks or months due to bridge structural failure. Spatial accessibility is measured in terms of accessibility scores, i.e., the accessibility score is calculated for each simulation and the mean value is presented, where the bigger the score the better the accessibility.

**Study Area and Data Sources**

The U.S. Gulf Coast Region has experienced major hurricanes and severe damages from past events (Ray et al. 2011), motivating a need to understand issues such as post-event accessibility to support future mitigation planning. The precise focus area of this study is the eastern part of Harris County, Texas, which is bounded by the Galveston Bay to the southeast and is an area vulnerable to storm surge and waves (Torres et al. 2017). In the past, major hurricanes have impacted the area. For example, Hurricane Ike in 2008 produced the highest recorded storm surges in Texas (Sebastian et al. 2014) and caused significant damages to the infrastructure network associated with storm surge and waves (Christian et al. 2014). This setting makes the area's transportation network particularly vulnerable to inundation and structural failures associated with heavy rainfall, storm surge and waves.

      The data sources for the analysis include locations and characteristics of supply facilities, road infrastructure and hurricane hazards. Health services facilities, including hospitals and primary care facilities, were obtained from ReferenceUSA (http://www.referenceusa.com) and geocoded into GIS formats. Population by census block groups (breakdowns by socioeconomic groups) is used as demand and the GIS boundary files for census block groups were downloaded from the U.S. Census Bureau TIGER/Line Products. The road network in GIS format is also obtained from the TIGER/Line Products and includes interstate, highway, arterials and local streets.

The locations and attributes of 88 bridges were gathered as a subset of those from Ataei and Padgett (2012), based on National Bridge Inventory (NBI 2017) data along with detailed plan review. As shown in Fig. 6, the study area consists of 1,021 healthcare facilities (supply), 121 census block groups (demand), 88 bridges and 3996 km of roadway (infrastructure). Two clusters of healthcare services emerge—one in the eastern part of the study area and the other in the southwestern corner. Two synthetic storm scenarios are considered in this study in order to determine surge and wave characteristics for the study region. Storm surge elevations and wave heights for these two scenarios are provided by the SSPEED Center (SSPEED 2017). The two scenario events, referred to herein as Storm 1 and Storm 2, are a subset of the synthetic events produced by FEMA (Douglass et al. 2004) as a part of its Flood Insurance Study, with surge elevations corresponding to approximately 100- and 500-year return period events (Bernier et al. 2017). The surge levels in the Houston Ship, an area enveloped by the study area, vary between 5.3 and 6.5 m for Storm 1 and between 6.5 and 7.8 m for Storm 2. The extent of inundation and the bridge uplift probability for the study area are shown in Fig. 7 for Storm 1 and in Fig. 8 for Storm 2. Based on this uplift probability, 1,000 numbers of samples ($N = 1,000$) are generated for each bridge for the Monte Carlo simulation (MCS) network analysis, given that accessibility scores were found to converge for $N > 400$. Other storm surge models can be used to estimate storm surge characteristics (Helderop and Grubesic 2018), but the examination of different surge models is out of the scope of the current study.

**Data Processing and Model Parameters**
Spatial accessibility varies depending on the state of the bridge or roadway that residents have to cross in order to access the healthcare facility. For example, an inaccessible bridge due to deck inundation or unseating, will force residents to follow an alternative path to reach the facility of interest. Residents' spatial accessibility to healthcare facilities is measured through several components. A census block group's overall population is considered the demand capacity, and this capacity is located at the centroid of the census block group for measuring purposes. Due to data unavailability, the number of employees in health care facilities is used as a proxy of supply capacity. Finally, network travel time between block group centroids and healthcare facilities is used to quantify the travel cost between them. The catchment area size $d_0$ is set to be 50-minutes, as our traffic network analysis finds that all residents in the study area are covered within the service area of 50-minute travel time (free-flow travel time) from any health care facilities. In this context, the accessibility score $A_i$ from Eq. (7) is essentially a supply-to-demand ratio (the ratio of the number of employees at healthcare facilities to that of block group population) with only selected supply locations (healthcare facilities) and demand locations (block groups) entering the numerator and the denominator, respectively. The selections are based on a 50-minute travel time threshold within which supplies and demands interact. For example, when selecting relevant demand sites to a supply location, only the ones falling inside the 50-minute travel time catchment area of the supply location are used. As the supply is oftentimes far less than the demand, accessibility scores can be far less than one. Thus, in order to avoid small values, the accessibility scores were uniformly inflated by multiplying all values by 1,000. The revised scores may be interpreted as the number of employees in health care facilities per 1,000 residents. Again, a bigger accessibility score denotes better access to facilities. The accessibility scores can be used to inform changes in the accessibility of residents between different storms and/or different time horizons. For example, how and to what extent futures storms are expected to affect resident's accessibility in order to formulate mitigation plans. However, these accessibility values can be affected by the choices of several parameters, such as the catchment area size and the definition of supply capacity. Thus, it can be difficult to find a clear-cut threshold of accessibility value that will target vulnerable neighborhoods and inform mitigation plans. Therefore, this study focuses on interpreting the changes in accessibility scores between two storm scenarios.

**Accessibility Analyses**
The analyses for the short- and long-term time horizons are presented for each storm scenario, resulting in four accessibility maps. Short-term effects refer to spatial accessibility measures considering bridge structural failure, bridge inundation, and road inundation. Both bridge and road inundation are determined given a deterministic safety threshold and bridge failure is determined based on the fragility model. Spatial accessibility scores for the short-term effects for Storm 1 and Storm 2 are reported in Fig. 9 and Fig. 10, respectively. To account for the variation of the accessibility score range between the storms, the block groups are grouped in each storm scenario by their accessibility score percentiles such as 25, 50, 75, and 100. Thus, for consistency, the four accessibility groups in both storm scenarios were rendered using the same color ramp, i.e., coral, yellow, cyan and purple.

In general, Storm 2 affects spatial accessibility to more areas than Storm 1, i.e., spatial accessibility to health services is reduced during stronger storms, as expected. For Storm 1, residents living in the southwestern part of the study area enjoy the best spatial accessibility to healthcare services (Fig. 9). This pattern is logical as most healthcare services are located in this area. However, for Storm 2, residents in the southwestern part are significantly affected by the inundation and bridge damage that occurs (Fig. 8). Thus, the actual landfall and storm characteristics can greatly affect residents' accessibility in areas that performed adequately during past events.

Long-term effects refer to spatial accessibility measures considering bridge structural failure, i.e., deck unseating based on the parameterized fragility, since they require longer time horizons for repair or replacement. Spatial accessibility scores for the long-term effects for Storm 1 and Storm 2 are reported in Fig. 11 and Fig. 12, respectively. Contrary to the short-term effects, all areas have at least some access to health services because inundation, which was the driver of short-term accessibility loss, is no longer an issue. Uplift probability will increase for some of the examined bridges between the two storm scenarios (Fig. 7, Fig. 8). Thus, comparison of Fig. 11 and Fig. 12 indicates a small change in the accessibility scores between the two examined storms and shows that only six census blocks will experience reduced accessibility. This is also related to the fact that the examined bridge network is not uniformly distributed in the study area.

The results reveal an overall improvement in accessibility over time. However, they also suggest that the census blocks of concern shift over time as well. For Storm 1, the most critical blocks in terms of accessibility shift from no access (Fig. 9) to having access (Fig. 11). However, there are blocks in the southwestern part of the study area that experience reduced accesibility over time. Thus, as the storm intensity increases and bridge damage becomes a more significant driver of loss of access, the census blocks affected by those routes with damaged bridges remain a concern for both short- and long-term time horizons. Although outside the scope of this analysis, in such a scenario repair/replacement resources and recovery activities are essential to affecting long-term accessibility and ultimate resilience.

The average coefficient of variation (COV) of the accessibility scores is generally low for each storm scenario and for both time horizons (long- and short-term). For example, for the long-term horizon the average COV is 0.0019% and 1.26% for storm 1 and storm 2, respectively. The close to zero COV indicates that storm 1 affects a few bridges which is the only source of uncertainty considered (i.e., bridge damage only through the uplift failure probability). Then, the average COV slightly increases for storm 2 as a more intense storm is expected to affect more bridges in the same study region. Thus, COV changes can be informative for sources of uncertainty that play a critical role to accessibility.

**Sociodemographic Assessment**
Sociodemographic maps, plotted from data which contain information on population characteristics such as age and household income are compared to the derived accessibility maps. This comparison can provide information about how different sociodemographic groups are affected by a hazard for shaping future policy/mitigation strategies. This study used sociodemographic indicators obtained from the US Census Bureau (https://www.census.gov) to look at the total population, age and household income rates in the study area. The total population map (Fig. 13) shows a significant portion of people located in the northwestern part of the study area who are expected to experience low accessibility for both time horizons of Storm 2. People over 65 years old and those with a household income below the poverty level are vulnerable groups that may not be easily evacuated ahead of coastal storms. The results reveal that people over 65 (Fig. 14) located in the central region of the study area will have adequate access to health services. However, comparison of Fig. 14 and Fig. 10 shows that the part of this population located in the southwestern part of the study area may experience no access at all in the case of an intense storm such as Storm 2. In addition, comparison of Fig. 11, Fig. 12 and Fig. 14 reveals that the part of this population located in the northeastern part of the study area may experience a decreased long-term accessibility to health services in both examined storms. Households with an income below the poverty level (Fig. 15) are mostly located in the central-east and central-west parts of the study area. Bridge structural failures do not seem to affect these areas in the long-term in both examined storms. However, the central-east area may experience a loss of short-term access.

To better understand the accessibility patterns among the selected sociodemographic groups (people over 65, people below poverty line, people above poverty line), the weighted average accessibility values were computed for each group to measure their health care accessibility. The weighted average accessibility is formulated as follows (Ikram et al. 2015):

$$\bar{A} = \frac{\sum_{i=1}^{n} p_i A_i}{\sum_{i=1}^{n} p_i} \tag{8}$$

where $p_i$ represents the population of a specific sociodemographic group in block group i. The results are reported for long term (Fig. 16) and short term (Fig. 17) horizons. The weighted average value for the overall population is also provided in Fig. 16 as a baseline for comparison. As shown in Fig. 16, compared to the general population, the below-poverty group on average had better spatial access to health care services in terms of long term effects for both storms, while the above-poverty group on average had more limited health care accessibility. Again, this pattern is largely attributable to the concentration of the below-poverty group in central part of the study area, where the impacts of bridge structural failures were minor. This can also explain the relatively greater health care accessibility of the senior population group, i.e., over 65 years old (Fig. 14) over the general population (Fig. 13). As for short term effects, a consistent pattern was present for storm 1. However, the pattern was exactly reversed for storm 2 (Fig. 17). When the storm becomes stronger, on average, health care accessibility declined for both below-poverty and senior population groups, whereas it improved for the above-poverty group. In addition to bridge structure failure, the short term effects also account for inundation. As illustrated in Fig. 14 and Fig. 15, elderly and poorer populations are more concentrated in central and southwestern parts of the study area. Though less impacted by bridge structural failure, these areas are more likely to be inundated, especially during a stronger storm such as storm 2 (Fig. 10). As a number of block groups had no access at all during either storm—for example, 18% of block groups in the study area lacked access to health care services during storm 1, while 40% of block groups during storm 2—the comparison of the average accessibility score between the short and long term effects would be biased and hence is not discussed.

**Discussion and Conclusions**
This study addresses a current gap in modeling and understanding limited spatial accessibility to critical services caused by coastal hazards. A framework that combines infrastructure vulnerability modeling and spatial accessibility modeling is presented with an application to an area prone to storm surge and waves caused by hurricanes. The integrated framework fills current gaps by considering different closure conditions in the aftermath of a coastal hazard, and it is flexible to account for deterministic and probabilistic models of infrastructure performance. Furthermore it leverages GIS models coupled with infrastructure performance data to evaluate accessibility to critical facilities, like healthcare, using a spatial accessibility model. The illustrations posed here emphasize the integration of spatial and temporally varying transportation infrastructure availability due to different closure conditions following storm events. In the future this model could also be used to consider post-hazard variations in supplies and demands regarding the service of interest (e.g. healthcare).

The case study of southeastern Harris County, Texas, illustrates the utility of the framework and insights gained. Two hypothetical storm scenarios are selected and two time horizons are examined for each hypothetical storm. The results indicate changes in the accessibility scores depending on the time horizon and storm intensity. Areas with good long-term access to health services, i.e., no bridge structural failures, can lose short-term access mainly due to flooding. However, the results indicate that spatial accessibility can also be highly related to the infrastructure vulnerability alone. By coupling the post-hazard spatial accessibility framework with sociodemographic analyses, new insights emerge regarding the populations most effected by the event. The case study revealed that vulnerable populations, such as those with low incomes or over the age of 65, are likely to have low accessibility to health services immediately following even a low level storm. The populations most effected in the long-term tend to be people with low incomes over the age of 65 who live in the northern part of the study area. While the current spatial accessibility analysis is conducted at the aggregated block group level, it can also yield insight on potentially affected populations when integrated with sociodemographic analysis, as demonstrated in this paper.

Although, the presented study offers advances in posing an integrated framework for assessing post-hazard accessibility, there are several assumptions and limitations of the present application that can serve as a launching point for future studies. For example, this paper presents mean estimates of spatial accessibility for scenario storm events. Future work can more fully characterize the probability distribution of these accessibility metrics for both scenario and probabilistic hazard simulations. Bridge damage is related only to the uplift failure of the deck. However, the infrastructure vulnerability modeling can be expanded to consider additional failure modes, such as debris related issues, in order to capture the explicit time evolution of restoration of functionality. Additionally the accessibility analysis can incorporate loss of supply, such as damage to health care facilities, or be expanded to evaluate access to other critical facilities and services.

Future work will address the need for fragility models parameterized on other bridge characteristics (e.g., number of girders, slope of deck, etc.), to accommodate vulnerability assessment for different types of superstructures within the same bridge and partially correlated span failures, as well as to explore sensitivity of the accessibility scores given different fragility models. Furthermore, alternative hazard and climate conditions should be explored and the framework could be used to evaluate the effects of mitigation and adaptation strategies on short and long term accessibility.

**Acknowledgements**

The authors grateful acknowledge the support of this research by the Shell Center for Sustainability. Partial support was also provided by the National Science Foundation through Award OISE-1545837. Any opinions, findings, and conclusions or recommendations expressed in this material are those of the authors and do not necessarily reflect the views of the sponsors. The authors also thankfully acknowledge Dr. Benjamin Bass and the Severe Storm Prediction, Education and Evacuation from Disasters (SSPEED) Center for providing the coupled ADCIRC and SWAN results for the examined storm scenarios.

**Notation**

The following symbols are used in this paper:
$A_i$ = accessibility;

$$\begin{aligned}
\bar{A} &= \text{Weighted average accessibility;} \\
a &= \text{regression coefficient;} \\
b &= \text{regression coefficient;} \\
c &= \text{regression coefficient;} \\
D_k &= \text{demand size;} \\
d_{in} &= \text{inundation depth;} \\
d_0 &= \text{catchment area around facility } j; \\
H_b &= \text{bridge deck height;} \\
H_{max} &= \text{maximum wave height;} \\
H_r &= \text{road surface height;} \\
H_s &= \text{significant wave height;} \\
H_{st} &= \text{storm surge height;} \\
i &= \text{demand location;} \\
j &= \text{facility location;} \\
k &= \text{location within the catchment area } d_0; \\
N &= \text{number of samples given } p_f; \\
n &= \text{number of block groups;} \\
p_f &= \text{probability of failure;} \\
p_i &= \text{population;} \\
R_j &= \text{supply-to-demand ratio;} \\
S_j &= \text{capacity of supply;} \\
Z_c &= \text{relative surge elevation;}
\end{aligned}$$

Table 1. Constants for calculating probability of failure adapted from Ataei and Padgett (2012).

| Mean span mass per unit length (ton/m) | $a$ | $b$ | $c$ |
|---|---|---|---|
| 0–5 | 0.6468 | 0.0406 | -0.1376 |
| 5–10 | 0.4166 | 0.0456 | -0.2343 |
| 10–15 | 0.3291 | 0.0546 | -0.2464 |
| 15–20 | −0.3300 | 0.0576 | -0.2444 |
| 20–25 | 0.2843 | 0.0512 | -0.2421 |
| 25–30 | 0.2865 | 0.0881 | -0.2391 |
| 30–35 | −0.1870 | 0.0782 | -0.2618 |

Note: if $a + bH_{max} + cZ_c < 0: p_f = 0$; if $a + bH_{max} + cZ_c > 1: p_f = 1$.

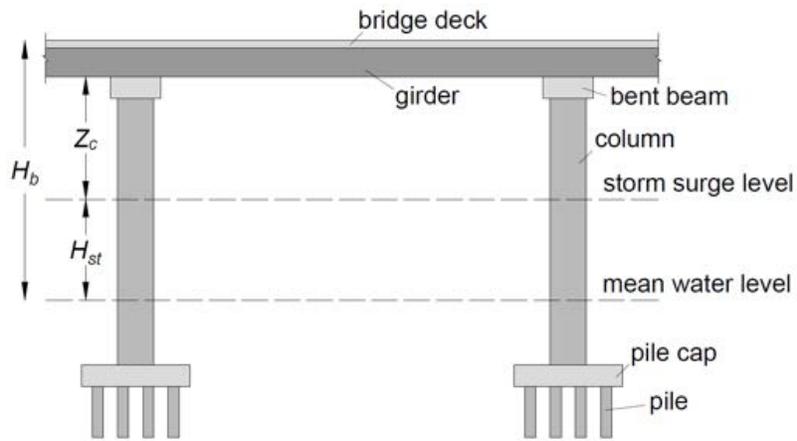

Fig. 1. Bridge and storm parameters

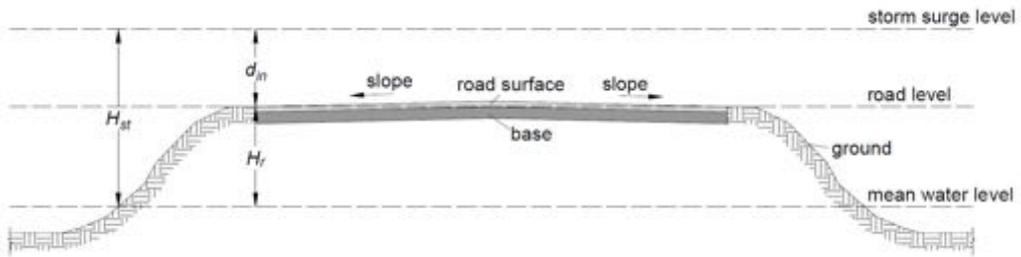

Fig. 2. Road and storm parameters

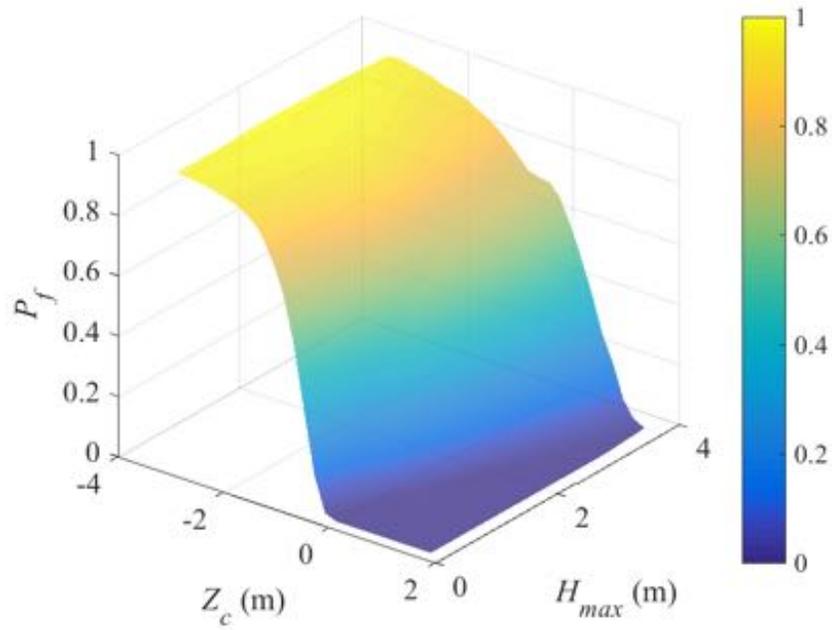

Fig. 3. Fragility surface for bridges with mean span mass of 15-20 ton/m

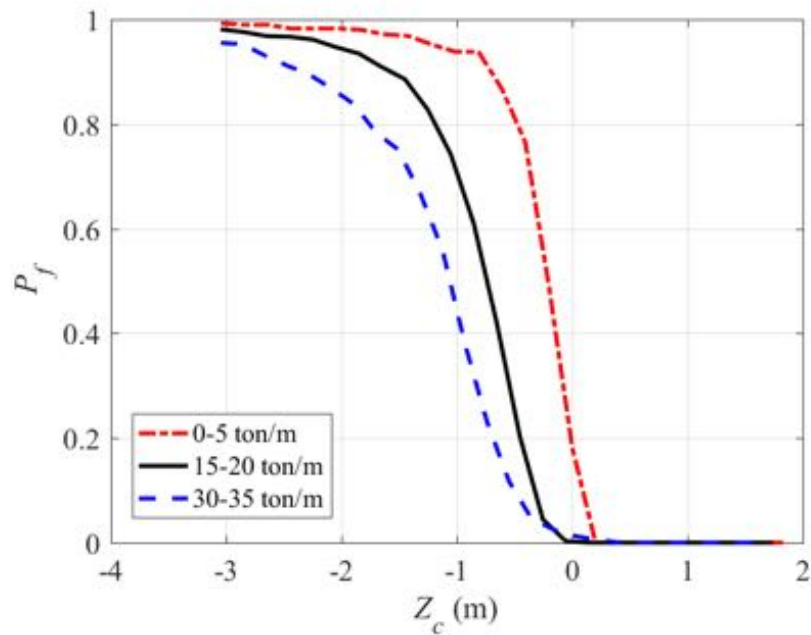

Fig. 4. Fragility curves for $H_{max}$=2 m and different mean span mass per unit length

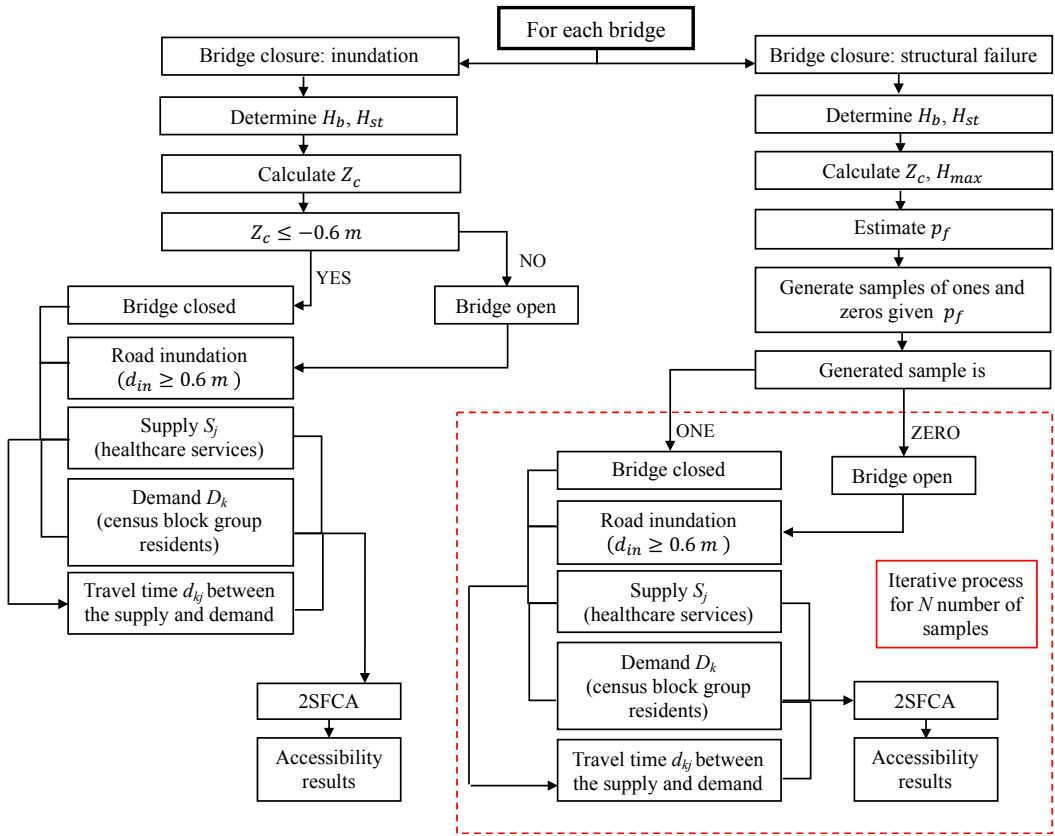

Fig. 5. Integrated framework of bridge vulnerability and spatial accessibility

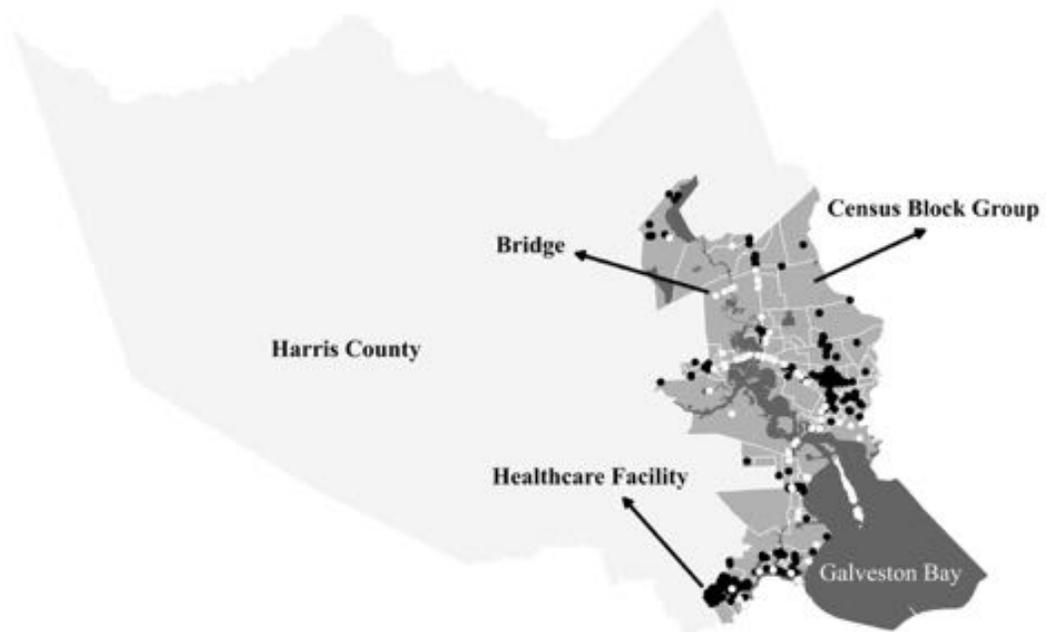

Fig. 6. Study area overlaid with locations of bridges and healthcare facilities

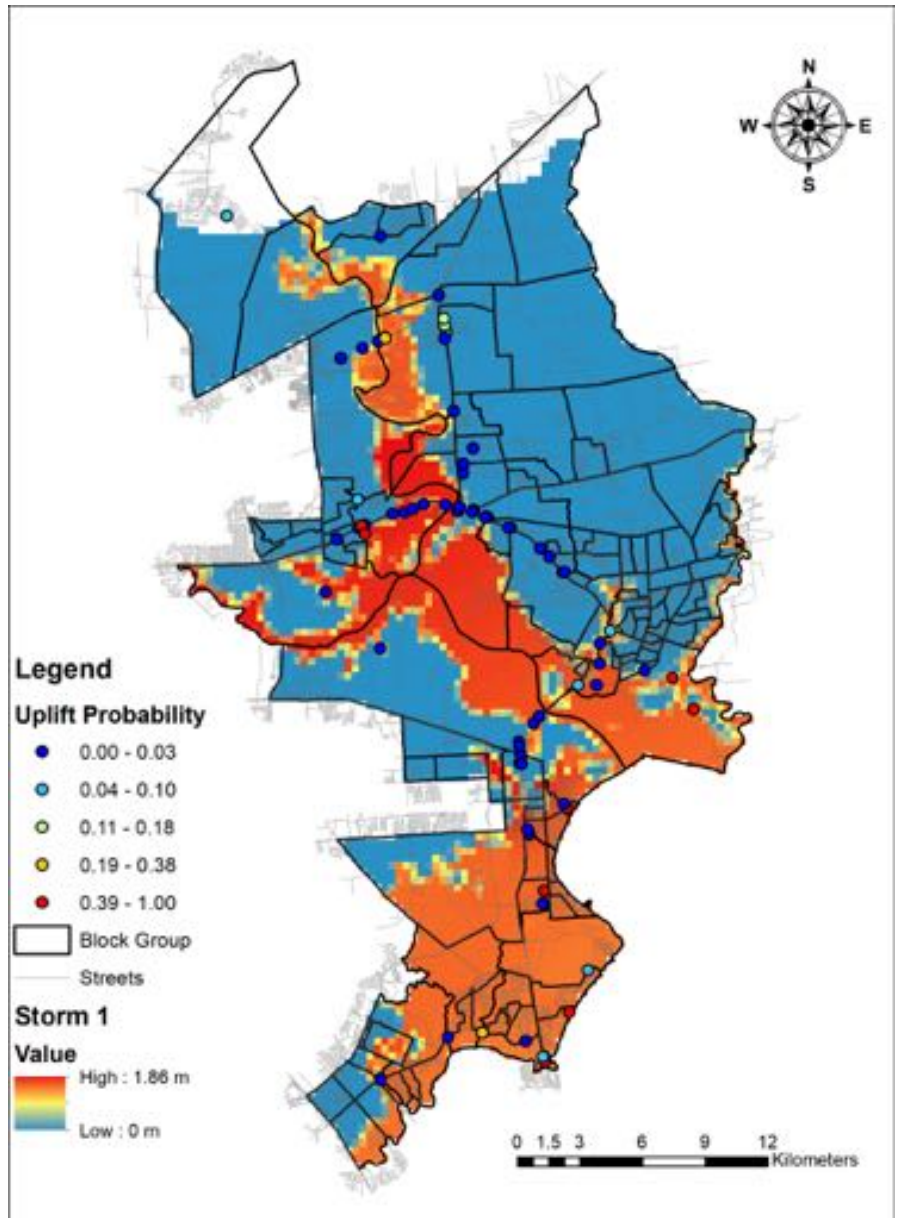

Fig. 7. Study area overlaid with inundation and bridge uplift probability for Storm 1

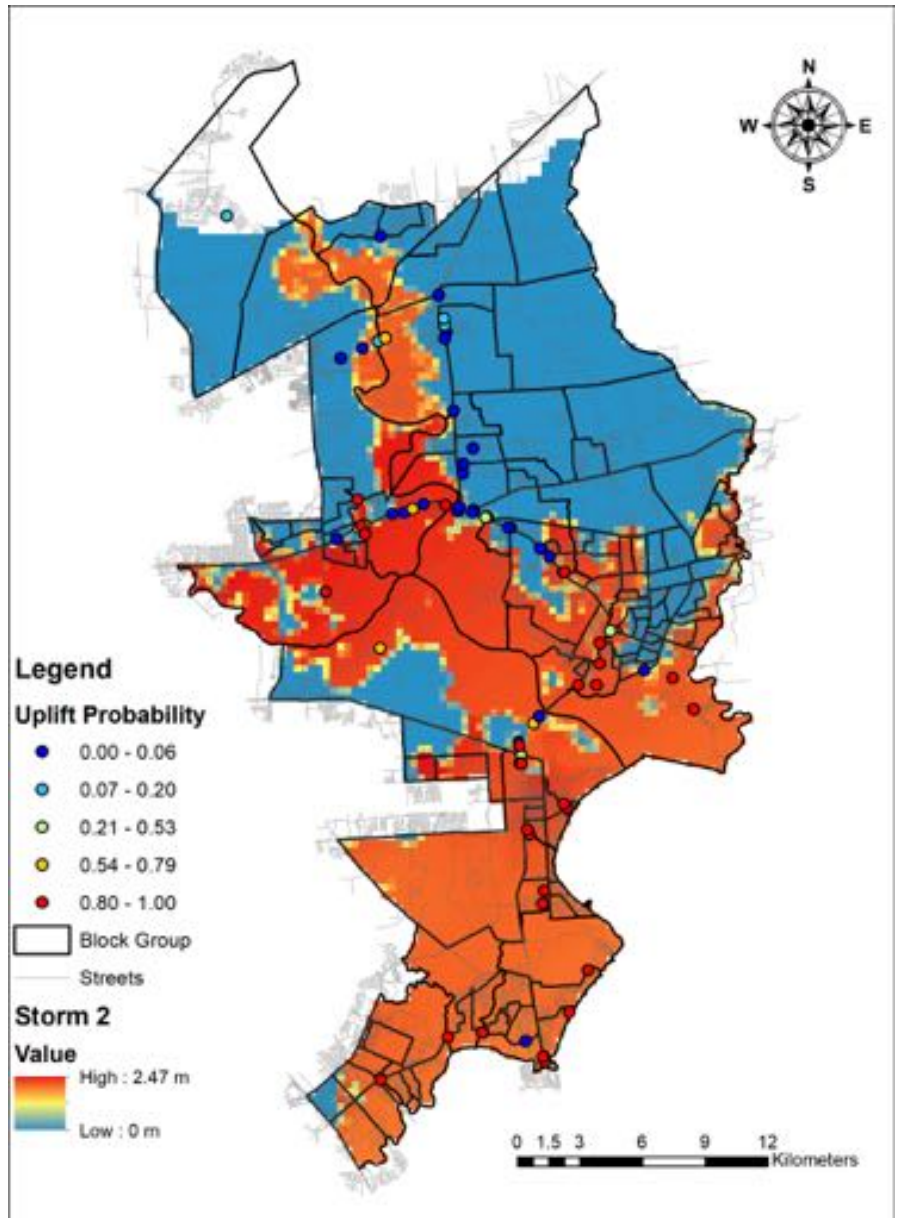

Fig. 8. Study area overlaid with inundation and bridge uplift probability for Storm 2

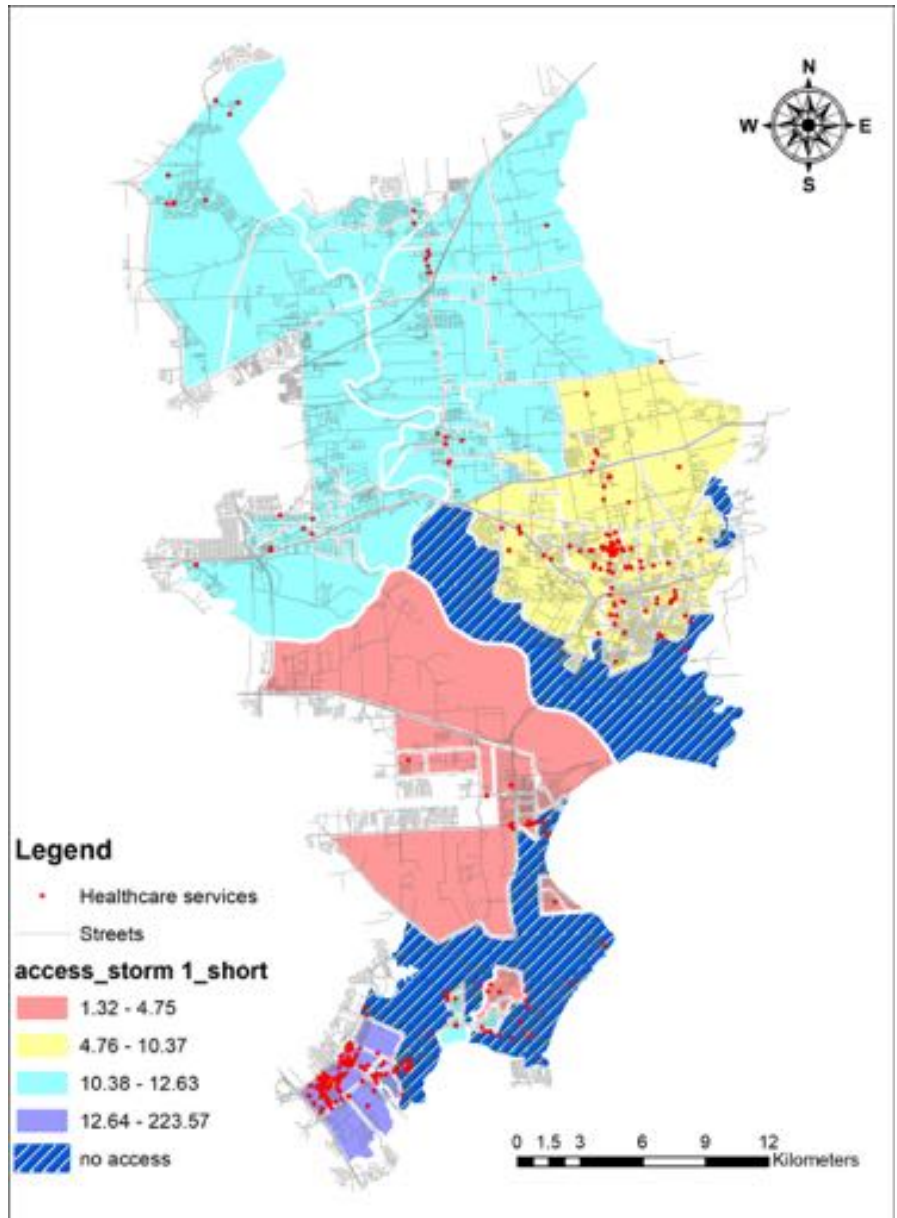

Fig. 9. Spatial accessibility scores for short term effects for Storm 1

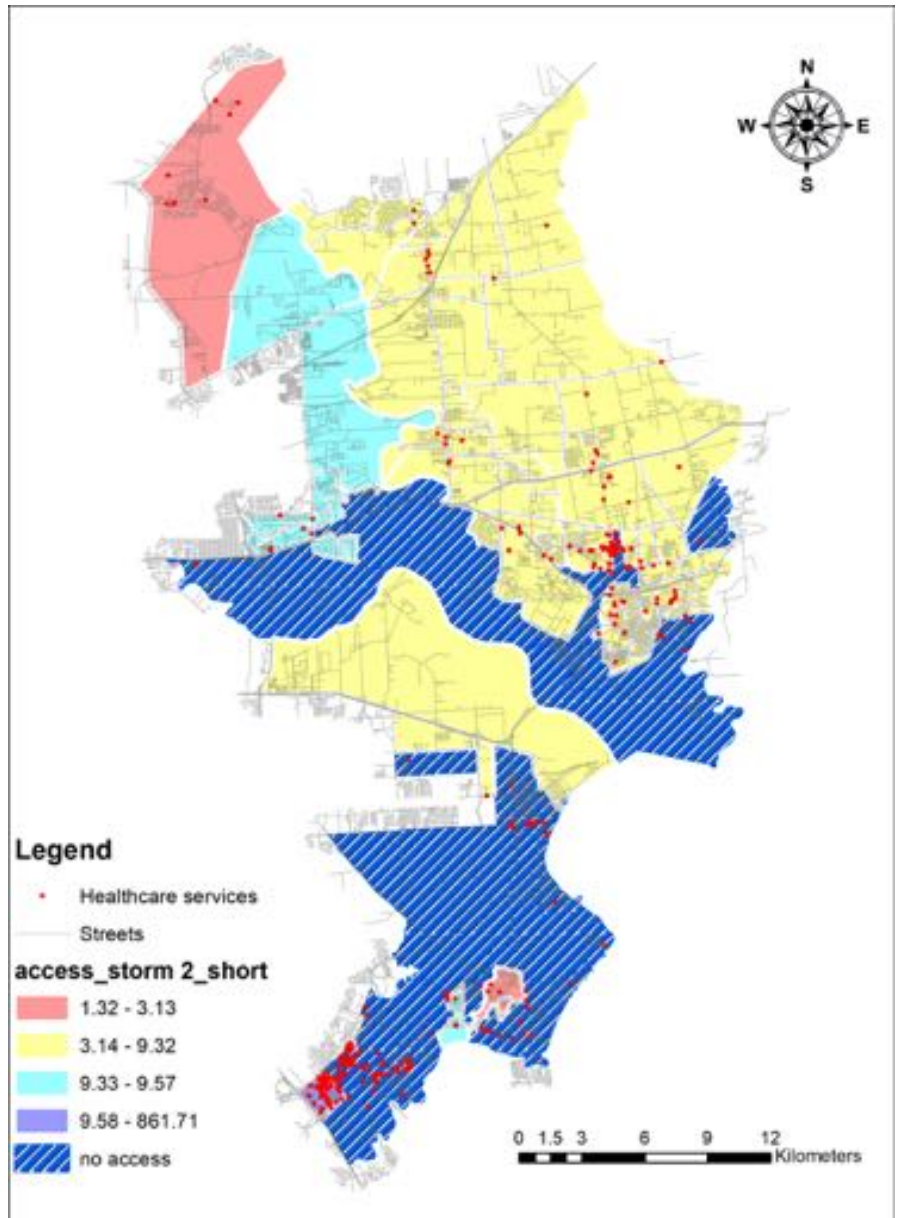

Fig. 10. Spatial Accessibility Scores for Short term Effects for Storm 2

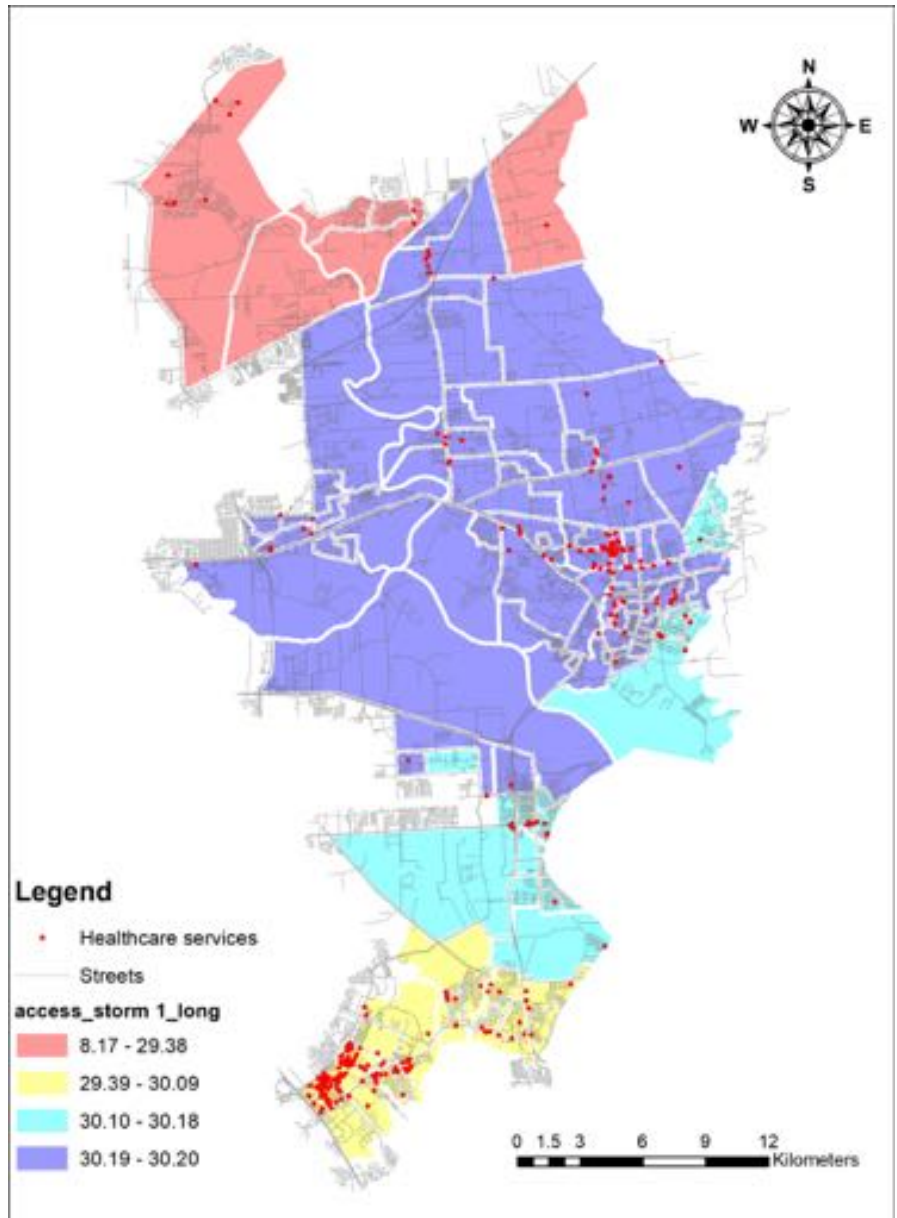

Fig. 11. Spatial Accessibility Scores for Long term Effects for Storm 1

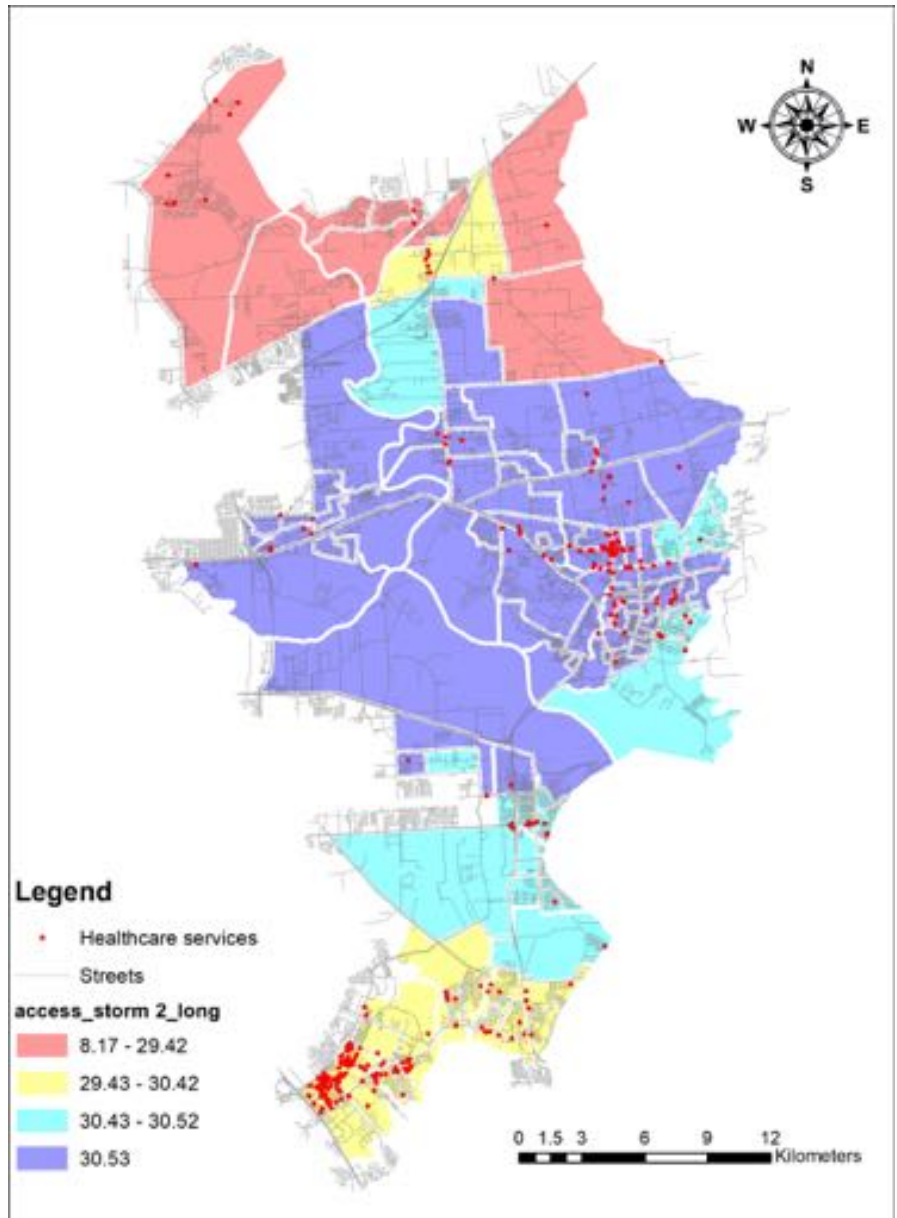

Fig. 12. Spatial Accessibility Scores for Long term Effects for Storm 2

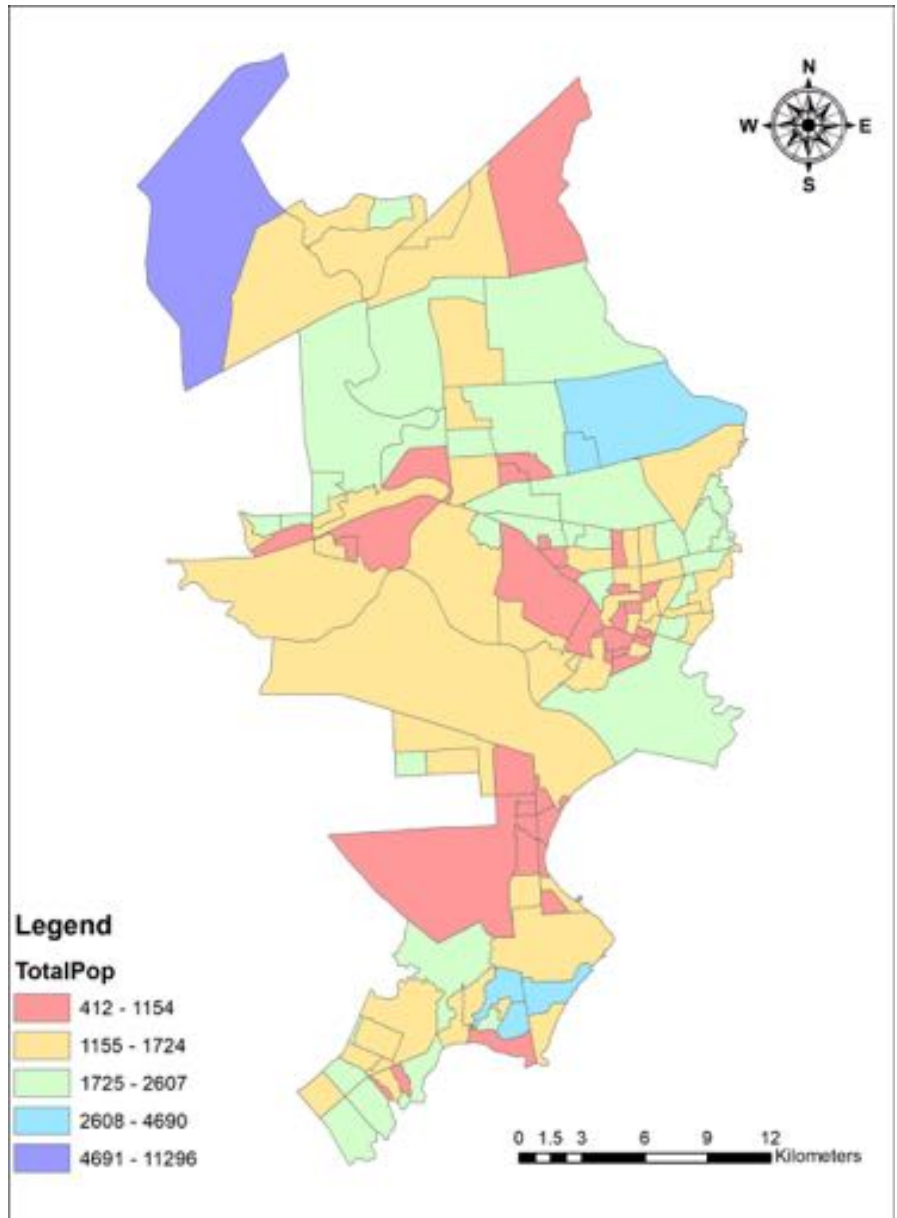

Fig. 13. Total population in number of people

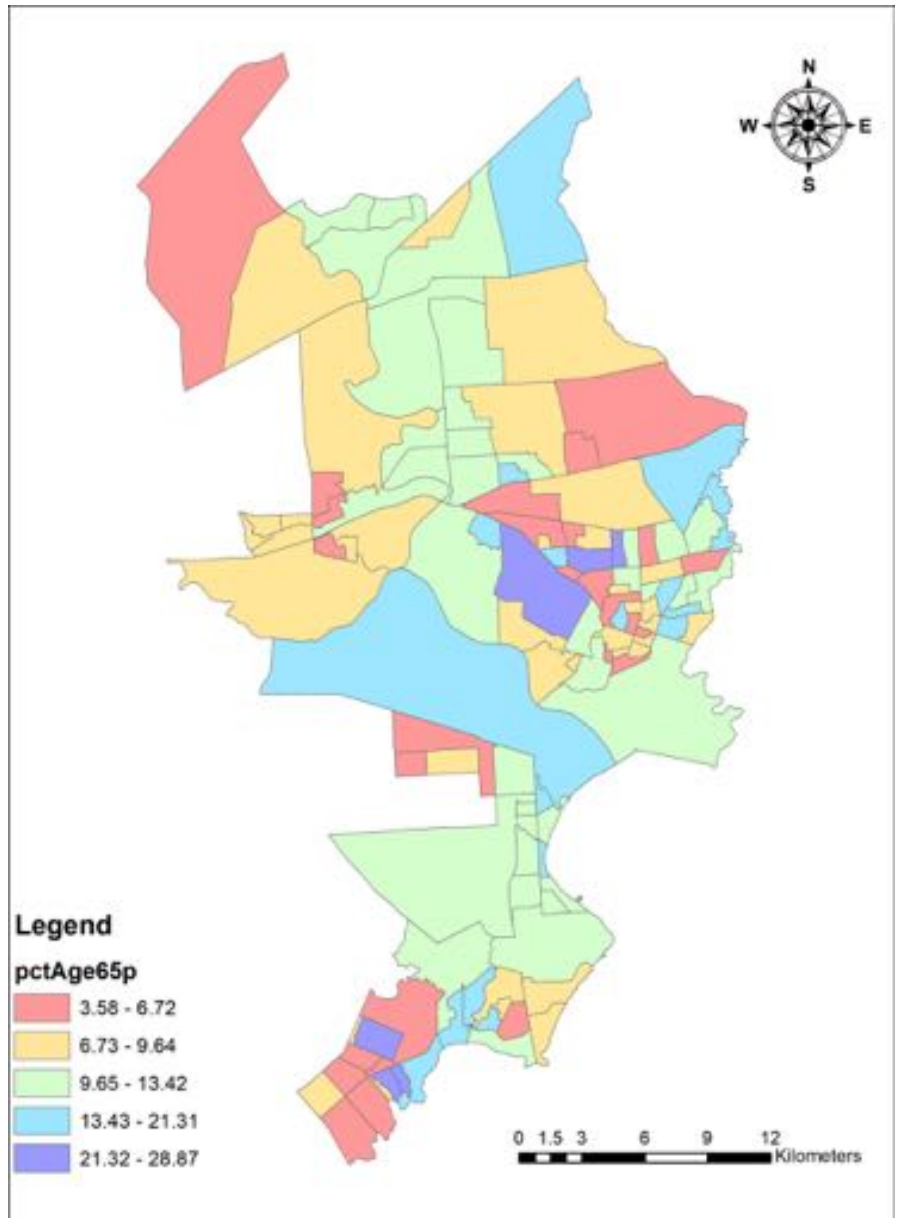

Fig. 14. Percentage population for age 65 and over

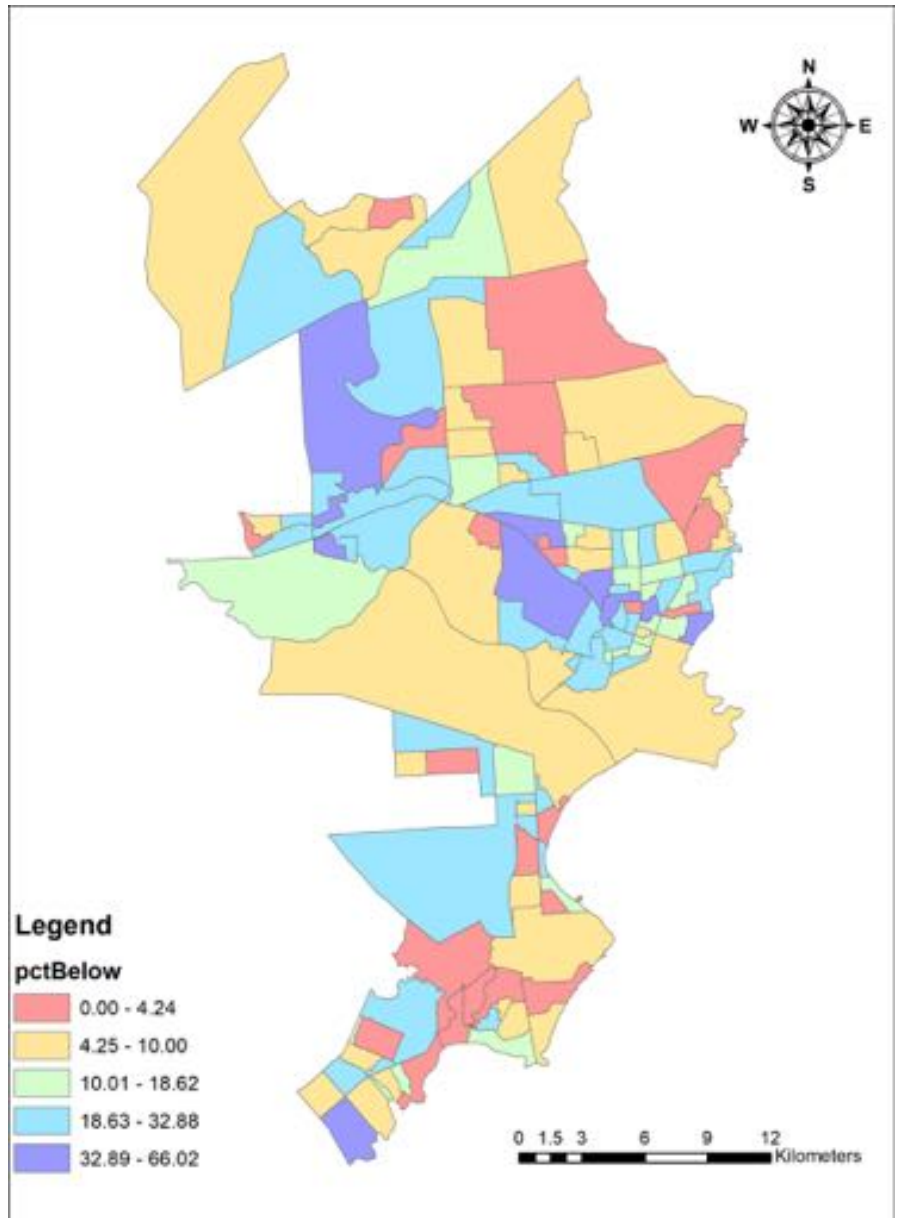

Fig. 15. Percentage household income below poverty level

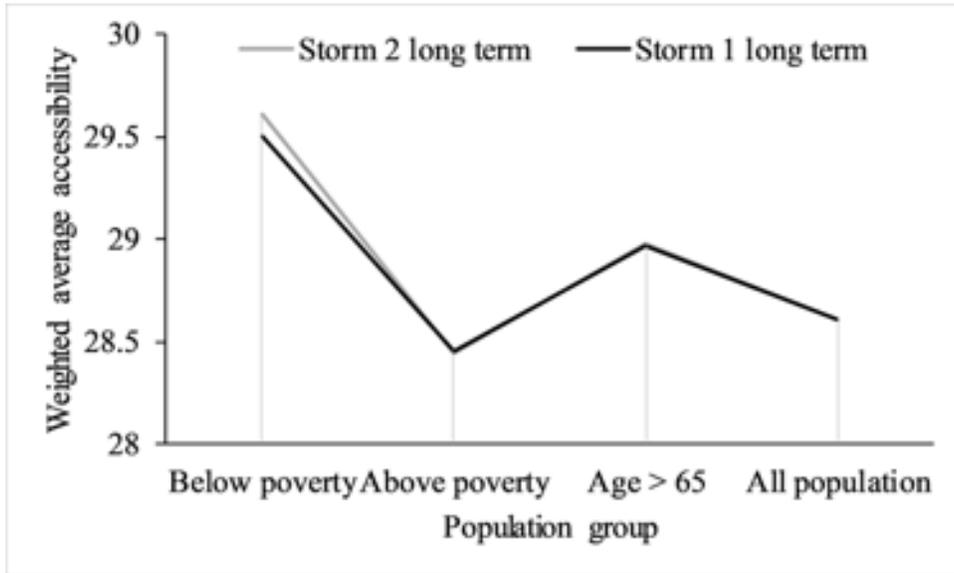

Fig. 16. Weighted average accessibility for sociodemographic groups under long term effects

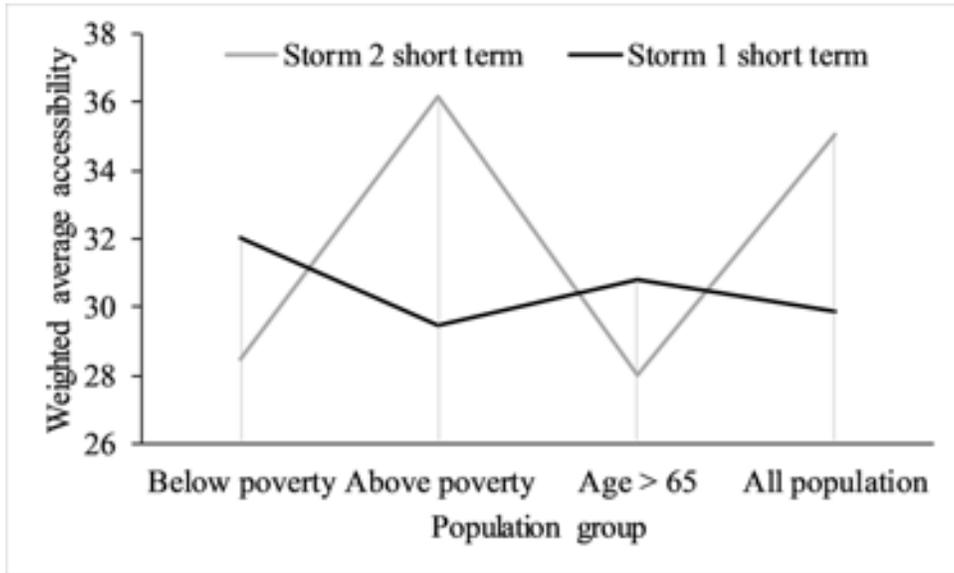

Fig. 17. Weighted average accessibility for sociodemographic groups under short term effects